%% file: fpcp11-abbott.tex
\documentclass[12pt]{article}
\usepackage{graphicx}
\newcommand\pubdate{\today}

\newcommand\pubnumber{}

\def\Title#1{\begin{center} {\Large #1 } \end{center}}
\def\Author#1{\begin{center}{ \sc #1} \end{center}}
\def\Address#1{\begin{center}{ \it #1} \end{center}}

\newcommand\pubblock{\rightline{\begin{tabular}{l} \pubnumber\\
         \pubdate  \end{tabular}}}
\newenvironment{Abstract}{\begin{center}{\bf Abstract}\end{center} \bigskip \begin{quotation}  }{\end{quotation}}
\newenvironment{Presented}{\begin{quotation} \begin{center} 
             PRESENTED AT\end{center}\bigskip 
      \begin{center}\begin{large}}{\end{large}\end{center} \end{quotation}}

\input econfmacros.tex
\begin{document}
\begin{titlepage}
\pubblock

\vfill

%%%%%%%%%%%%%%%%%%%%%%%%%%%%%%%%%%%%%%%%%%%%%%%%%%%%%%%
%%MODIFY
%%%%%%% TITLE, AUTHOR, ADDRESS 
%%%%%%%%%%%%%%%%%%%%%%%%%%%%%%%%%%%%%%%%%%%%%%%%%%%%%%%

\Title{CP Violation with $B_{s} \rightarrow J/\psi \phi$ at the Tevatron}
\vfill
\Author{Brad Abbott}  
\Address{University of Oklahoma, Norman, OK, 73019, USA}
\vfill

%%%%%%%%%%%%%%%%%%%%%%%%%%%%%%%%%%%%%%%%%%%%%%%%%%%%%%%
%%MODIFY
%%%%%%% Abstract
%%%%%%%%%%%%%%%%%%%%%%%%%%%%%%%%%%%%%%%%%%%%%%%%%%%%%%%

\begin{Abstract}
Recent results using $B_{s}^{0} \rightarrow J/\psi \phi$ decays for measuring
the CP violating phase, $\phi_{s}$, and the decay width difference for
the two mass eigenstates, $\Delta \Gamma_{s}$,
are presented from the CDF and D0 experiments at the Fermilab Tevatron collider.
\end{Abstract}

\vfill

\begin{Presented}
The Ninth International Conference on\\
Flavor Physics and CP Violation\\
(FPCP 2011)\\
Maale Hachamisha, Israel,  May 23--27, 2011
\end{Presented}
\vfill

\end{titlepage}
\def\thefootnote{\fnsymbol{footnote}}
\setcounter{footnote}{0}
%

%%%%%%%%%%%%%%%%%%%%%%%%%%%%%%%%%%%%%%%%%%%%%%%%%%%%%%%
%%%%%%% Article body
%%%%%%%%%%%%%%%%%%%%%%%%%%%%%%%%%%%%%%%%%%%%%%%%%%%%%%%

\section{Introduction}

In the Standard Model (SM) the heavy (H) and light (L) mass eigenstates in the $B_{s}^{0}$ system are predicted to have
a large decay width difference, $\Delta \Gamma_{s} = \Gamma_L - \Gamma_H$.  For $b \rightarrow c\bar{c} s$ decays, a CP-violating
phase appears due to the interference in the decay with and without mixing.  This CP violating phase, $\phi_{s} = -2\beta_{s}$, is
expected to be small and have a value $\phi_{s}$ = -0.038 $\pm$ 0.002 \cite{Lenz,Bona}.
Beyond the SM physics processes can alter this phase to $\phi_{s}= -2\beta_{s} + \phi_{s}^{\Delta}$ providing an unambiguous
measurement of new physics if the measured CP violating phase is large.

\section{Measurement using $B_{s} \rightarrow J/\psi$ decays}

The decay of $B_{s} \rightarrow J/\psi \phi $ involves the decay of a pseudo-scalar into two vector mesons. 
In the rest frame of the $B_{s}$
there are three possible angular momentum states L=0,1,2.  The final state
is an admixture of CP even (L=0,2) and CP-odd (L=1) states so an angular
analysis must be performed to disentangle the final states.  For this purpose,
the transversity basis \cite{trans} is used by both experiments as it gives direct access to the CP related quantities.  

In the transversity basis there are 3 independent complex amplitudes
$A_{0}$, $A_{\parallel}$, and $A_{\perp}$.   By measuring the angles $\theta$, $\phi$ and $\psi$ it is possible to disentangle the 3 final states and measure
the CP eigenstates.   The time dependent angular distribution is a function of various parameters:
$$\frac{d^{4}\Gamma(B_{s} \rightarrow J/\psi \phi)}{dt\;  d\cos \theta \; d\psi \; d\phi} = f(\phi_{s},\Delta m_{s}, 
\tau_{s}, \Delta \Gamma_s, A_{0}, A_{\parallel}, A_{\perp}, \delta_{\parallel}, \delta_{\perp}), $$
where $\delta_{\parallel}$ and $\delta_{\perp}$ are strong phases.  By fitting the time dependent angular
distribution, it is possible to extract the parameters of interest.

Flavor tagging the initial state is very useful in this measurement.
Without flavor tagging, there is a sign ambiguity on the weak phase for a 
given $\Delta \Gamma_{s}$.  Flavor tagging the initial state as a $B_{s}$ or a $\bar{B_{s}}$ allows the removal of 
this ambiguity and also provides an improved measurement of the parameters $\phi_{s}^{J/\psi \phi}$ and $\Delta \Gamma_{s}$.  

The first measurements by D0 and CDF were made with $\approx$ 1 fb$^{-1}$ of data 
\cite{d01fb,CDF1fb}.  
Both experiments performed untagged analyses and found results consistent
with the SM at $\sim$ 1.5 $\sigma$.  The D0 experiment measured
$|\phi_{s}^{J/\psi \phi}|$ = 0.79 $\pm$ 0.56 (stat)$_{- 0.01}^{+ 0.14}$ (syst.) and
$\Delta \Gamma_{s}$ = 0.17  $\pm$ 0.09(stat) $\pm$ 0.02(syst.).
CDF measured the probability of a deviation from the SM as large as was observed in the 
data at 15\%.

\begin{figure}[h!tb]
\centering
\includegraphics[width=0.6\textwidth]{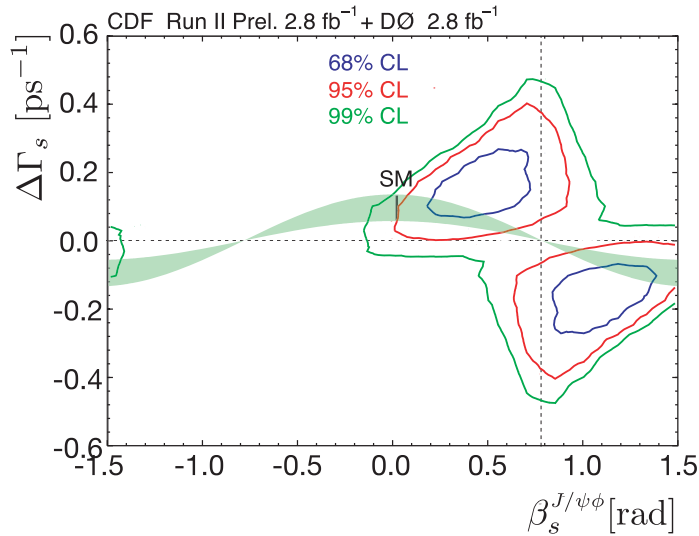}
\caption{Combined results from the D0 and CDF experiments on $\beta_{s}$ and
$\Delta \Gamma_{s}$ using 2.8 fb$^{-1}$ of data.}
\label{fig:comb28}
\end{figure}

With a larger data set of $\sim$ 2.8 fb$^{-1}$ both experiments performed
a flavor tagged analysis \cite{d028fb,cdf28fb}.  The results from both
experiments were consistent and both showed an approximately 1.8 $\sigma$ deviation 
from the SM expectations.  The two experiments combined their results
\cite{comb} giving a 2.1 $\sigma$ deviation from the SM expectation
(SM p-value=3.4\%), see Figure \ref{fig:comb28}.

\begin{figure}[htb]
\centering
\includegraphics[width=0.6\textwidth]{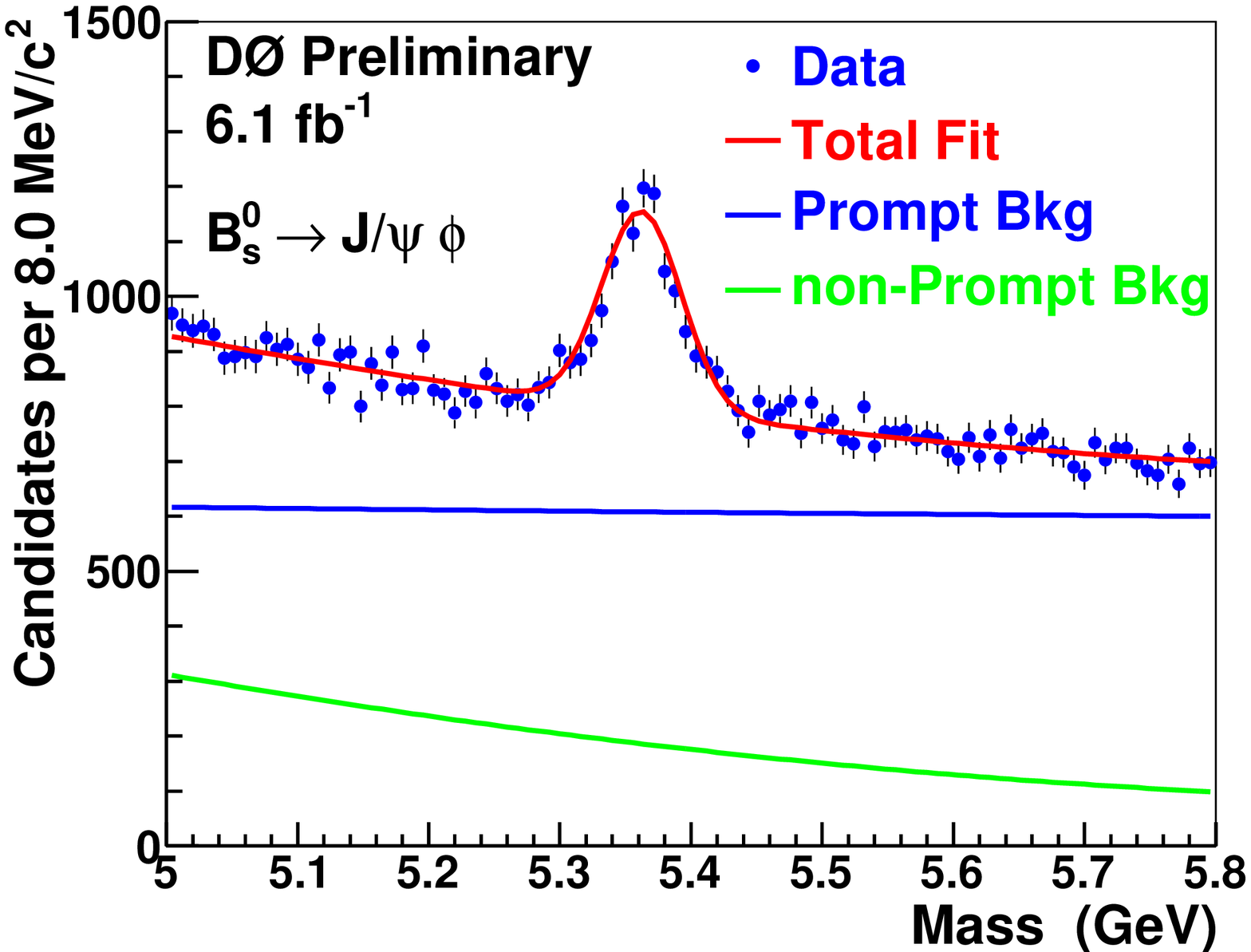}
\caption{The invariant mass of the $J/\psi \phi$ system for $B_{s}$ candidates from D0. A fit yields 3435 $\pm$ 85 $B_{s}$ candidates.}
\label{fig:bsyieldD0}
\end{figure}

\begin{figure}[Htb]
\centering
\includegraphics[width=0.6\textwidth]{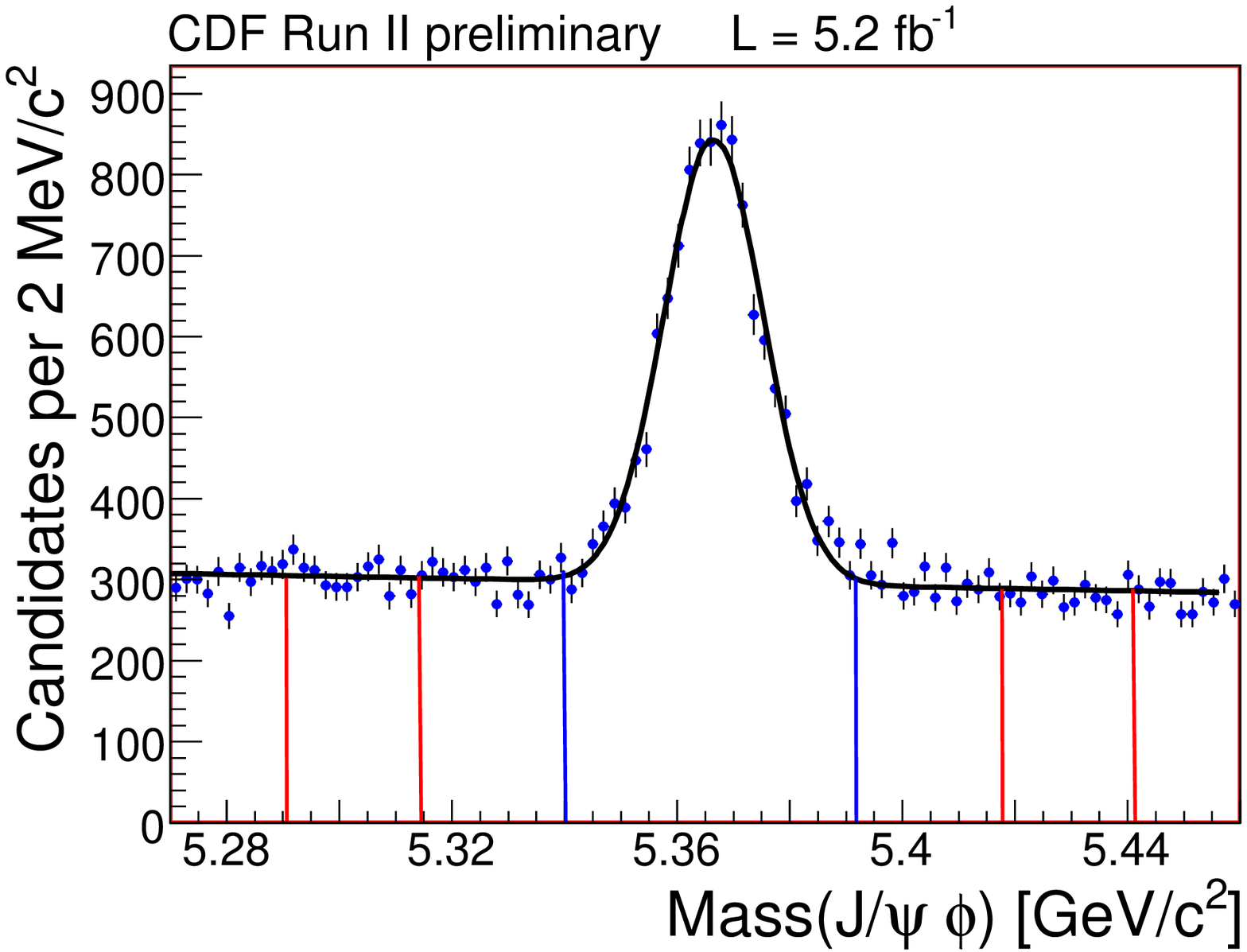}
\caption{The invariant mass of the $J/\psi \phi$ system for $B_{s}$ candidates from CDF. 
A fit yields 6405 $\pm$ 85 $B_{s}$ candidates}
\label{fig:bsyieldCDF}
\end{figure}

\section{Recent Results}

Larger data sets have allowed additional parameters to be extracted from the fits and have reduced
the uncertainties on the measurements.
In addition, improved event selection from D0 and improved particle ID, use of a neural net, 
improved same side flavor tagging and potential
S-wave contributions have been incorporated by CDF.  

Both experiments apply similar event selection based on kinematical
quantities (Transverse momentum of $B_{s}$, kaons and $\phi$) and both experiments
require good quality on all four tracks used to reconstruct the $B_{s}$,
and the four track vertex must have a minimum $\chi^{2}$ value.
In addition, CDF applies an additional neural net selection by 
combining particle ID with the kinematical variables.  
Figures ~\ref{fig:bsyieldD0} and \ref{fig:bsyieldCDF} show the yields from
D0 (6.1 fb$^{-1}$) and CDF (5.2 fb$^{-1}$).

Flavor tagging has also been improved from both experiments.  For opposite
side tagging, both experiments use a likelihood ratio or neural net as
a discriminating variable using an electron or muon from the opposite
side $B$ decay, the charge of the reconstructed opposite decay vertex
and opposite tracks charge (D0).  

The flavor tagging power can be characterized as $\epsilon D^{2}$ where
$\epsilon$ is the efficiency to provide a flavor tag and $D$ is the dilution defined 
as $D=\frac{\rm{correct} \; \rm{tags} - \rm{wrong} \; \rm{tags}}{\rm{correct} \; \rm{tags} + \rm{wrong} \;\rm{tags}}$.

\begin{figure}[htb]
\centering
\includegraphics[width=0.6\textwidth]{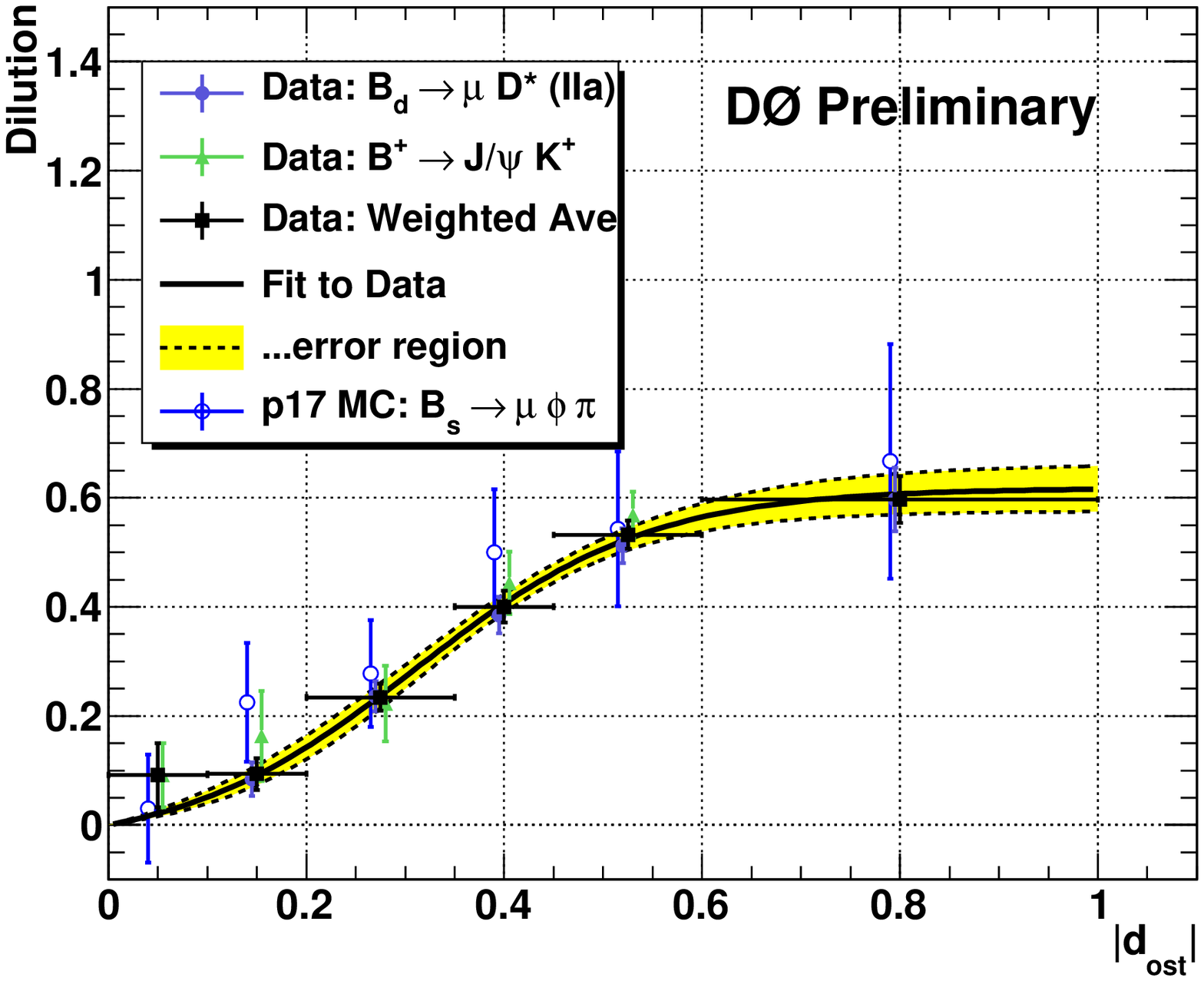}
\caption{Calibration of opposite side flavor tagging at D0.}
\label{fig:D0flavortag}
\end{figure}

\begin{figure}[htb]
\centering
\includegraphics[width=0.6\textwidth]{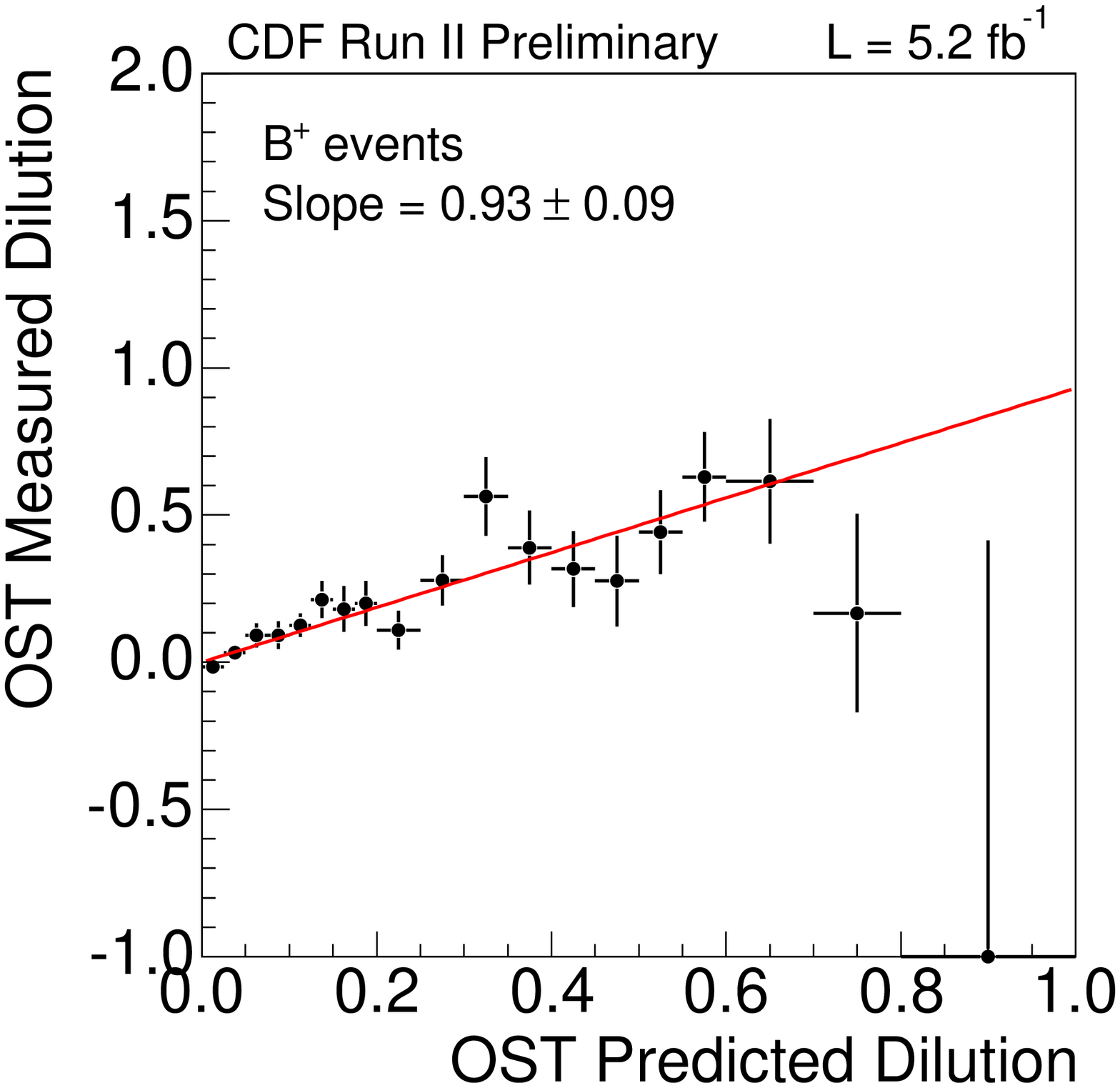}
\caption{Calibration of opposite side flavor tagging at CDF.}
\label{fig:CDFflavortag}
\end{figure}

Calibration of the opposite side tagging is performed using both
$B \rightarrow \mu \nu D^{* \pm}$ and $B \rightarrow J/\psi K^{\pm}$ decays, see Figures \ref{fig:D0flavortag} and \ref{fig:CDFflavortag}.

\begin{figure}[htb]
\centering
\includegraphics[width=0.6\textwidth]{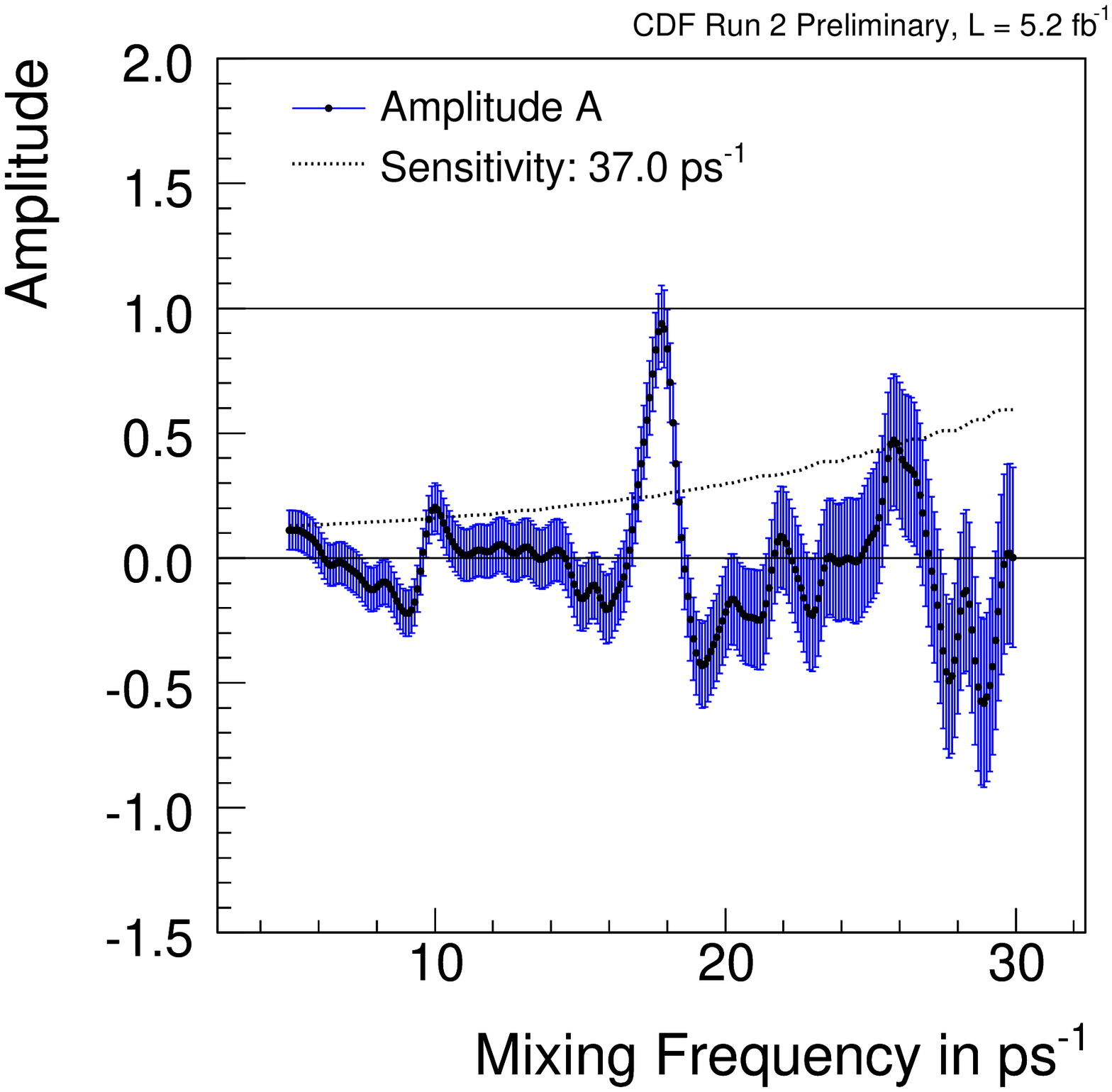}
\caption{Measurement of $\Delta m_{s}$ used to calibrate the same sign flavor tagger.}
\label{fig:cdfsstagger}
\end{figure}

In addition, CDF uses same sign tagging by using the sign of the charged
kaon produced during fragmentation.  Charged kaons are identified by using dE/dx information and the time
of flight.  The same sign kaon tagging was recalibrated for this
measurement by using $B_{s}$ mixing using several different decay modes.
The value of $\Delta m_{s}$ found from the calibration is 17.79 $\pm$ 0.07(stat) fb$^{-1}$, see Figure \ref{fig:cdfsstagger}, which is consistent with the PDG \cite{PDG}.
The measured amplitude is used to scale the same side tagger.

The opposite side tagging power measured at D0 is 2.5\%.  CDF measures 
an opposite side tagger power of 1.2\% and a same side tagging power of
3.1\%.  
During fitting, the measured value of the dilution is included on an event-by-event basis
as a tagging probability.

Possible S-wave contributions can arise due to $f_{0}$ or non-resonant
$K^{+}K^{-}$ production.  Both of these contributions give a flat distribution under the $\phi$
mass peak.  
CDF has measured the S-wave contribution, finding a value of 
less than 6.7\% for
1.009 $<$ $M_{K^{+}K^{-}}$ $<$ 1.028 GeV at the 95\% C.L.

\begin{figure}[htb]
\centering
\includegraphics[width=0.6\textwidth]{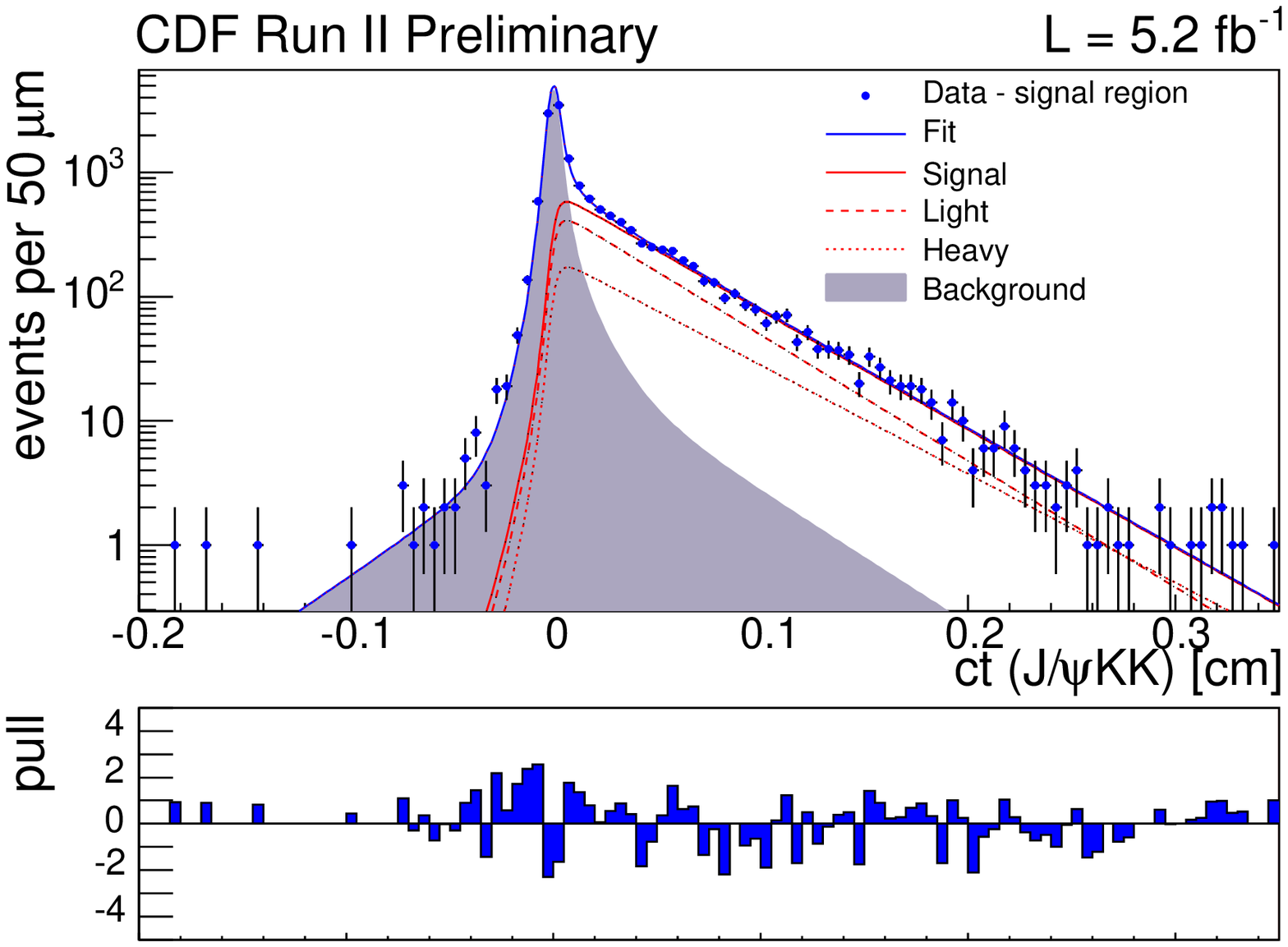}
\caption{Lifetime distribution of $J/\psi K^+K^-$ events at CDF.}
\label{fig:cdflifetime}
\end{figure}

The lifetime distribution obtained at CDF is shown in Figure \ref{fig:cdflifetime}.
Assuming no CP violation in the $B_{s}$ system, CDF obtains the
most precise single measurements of lifetime and decay width difference:
$\tau_{s}$ = 1.53 $\pm$ 0.025(stat) $\pm$ 0.012 (syst.) ps and
$\Delta \Gamma_{s}$ = 0.075 $\pm$ 0.025 (stat) $\pm$ 0.01 (syst.) ps$^{-1}$.

For D0 a multidimensional fit is performed applying a Gaussian constraint on
$\Delta m_{s}$ \cite{PDG} and on the strong phases $\delta_{i}$.  The values of
$\delta_{i}$ are obtained from $B^{0} \rightarrow J/\psi K^{*}$ decays under
a U(3) flavor asymmetry assumption.
D0 additionally assumes that the S-wave contribution is small and is not included
in the fit.

\begin{figure}[htb]
\centering
\includegraphics[width=0.6\textwidth]{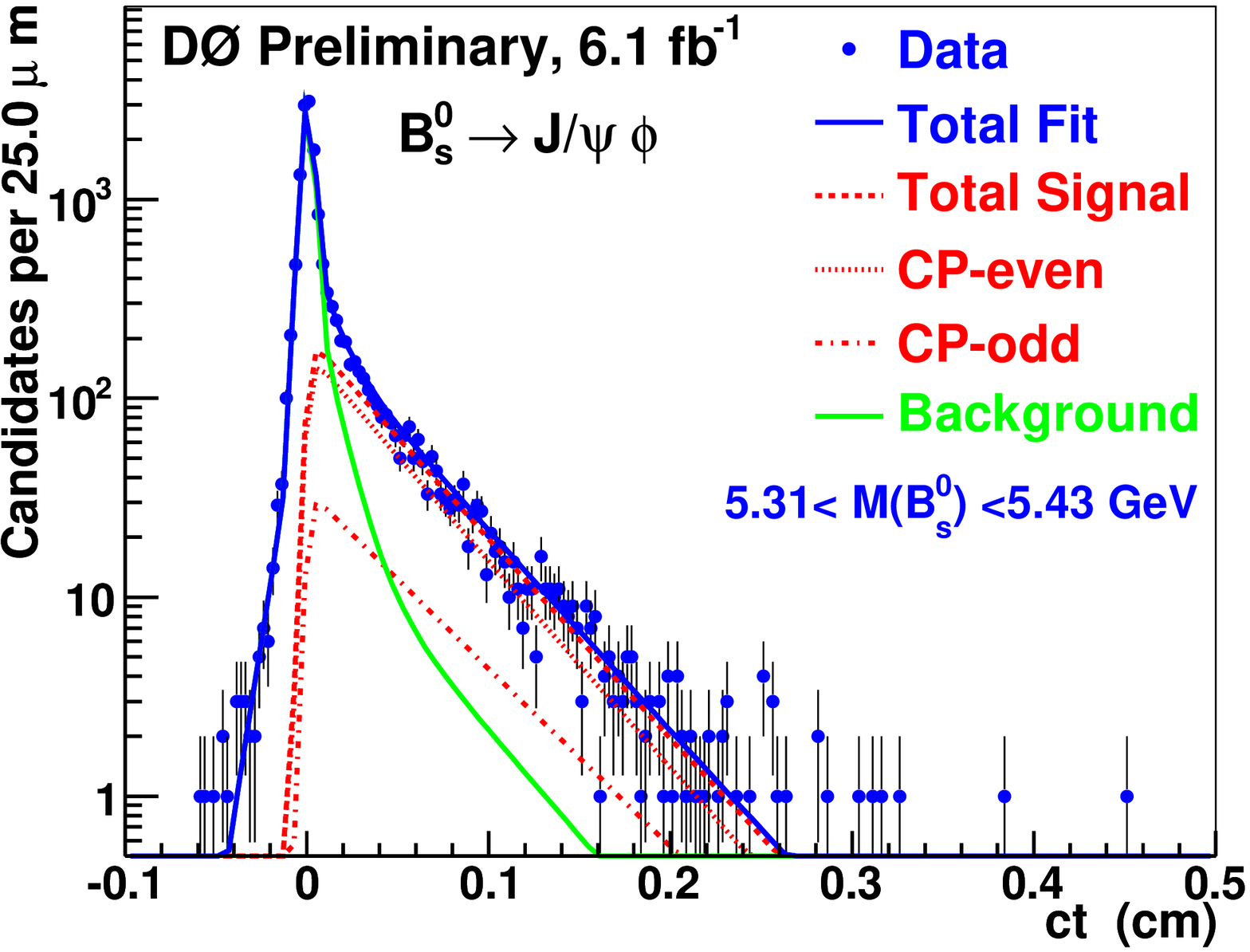}
\caption{Lifetime distribution of $J/\psi K^+K^-$ events from D0}
\label{fig:d0lifetime}
\end{figure}

Figure \ref{fig:d0lifetime} shows the lifetime distribution obtained at D0.  D0 obtains:
$\tau_{s}$ = 1.45 $\pm$ 0.04(stat) $\pm$ 0.01 (syst.) ps and $\Delta \Gamma_{s}$ = 0.15 $\pm$ 0.06(stat) $\pm$ 0.01(syst.) ps$^{-1}$
and $\phi_{s}^{J/\psi \phi}$ = -0.76 $\pm$  0.37 (stat) $\pm$ 0.02(syst.).

\begin{figure}[htb]
\centering
\includegraphics[width=0.6\textwidth]{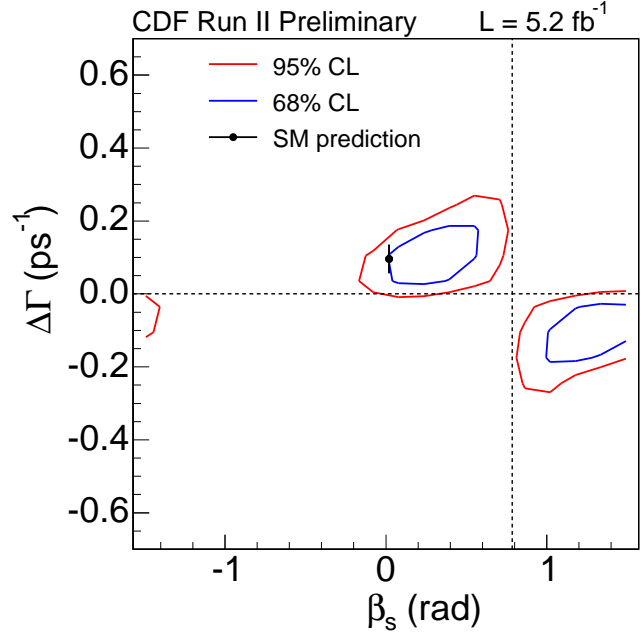}
\caption{68\% and 95\% C.L. contours in the plane $\Delta \Gamma_s$ - $\beta_{s}$  from CDF}
\label{fig:cdfcontour}
\end{figure}

\begin{figure}[htb]
\centering
\includegraphics[width=0.6\textwidth]{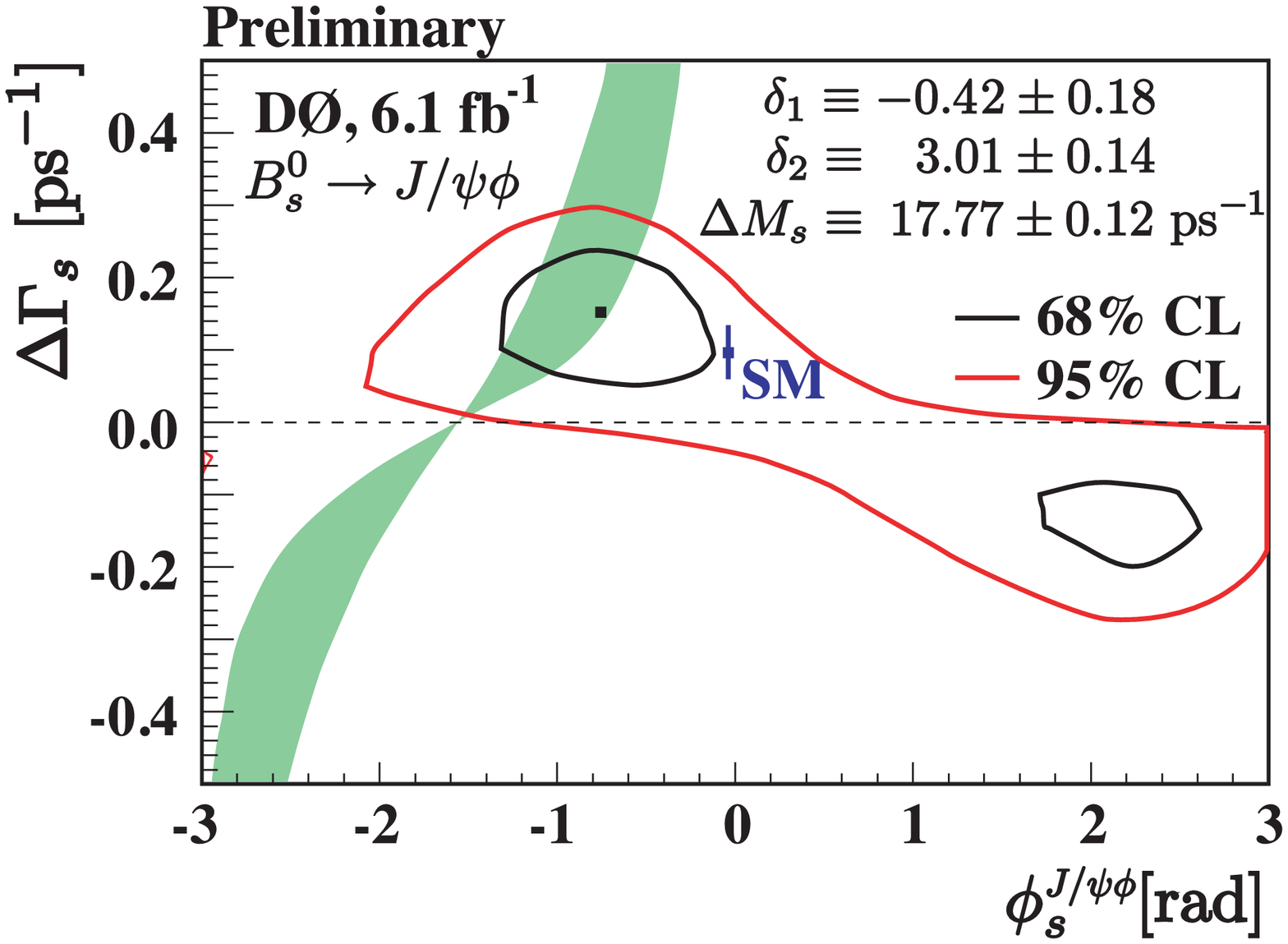}
\caption{68\% and 95\% C.L. contours in the plane $\Delta \Gamma_s$ - $\phi_{s}$ from D0}
\label{fig:d0contour}
\end{figure}

Contours in the $\phi_{s}^{J/\psi \phi}$-$\Delta \Gamma_{s}$ plane ($\beta_s$-$\Delta \Gamma_{s}$ plane) are generated for each experiment and are shown in
 Figures \ref{fig:cdfcontour} and \ref{fig:d0contour}.
D0 measures 0.014 $<$ $\Delta \Gamma_{s}$  $<$ .263 ps$^{-1}$ with
-1.65 $<$ $\phi_{s}^{J/\psi phi}$ $<$ 0.24 and
-0.235 $<$ $\Delta \Gamma_{s}$ $<$ -0.040 ps$^{-1}$ with
1.14 $<$ $\phi_{s}^{J/\psi phi}$ $<$ 2.93 at the 95\% C.L.
CDF finds $\beta_{s}$ $\in$ [0.02,0.52] $\bigcup$ [1.08,1.55] at the 68\% C.L.
The p-value of the SM central point = 44\% (0.8 $\sigma$ deviation from SM)

\section{Conclusion}

Using 5.2 fb$^{-1}$ and 6.1 fb$^{-1}$ of data respectively, 
the CDF and D0 experiments at the Tevatron have made the most direct and precise
experimental measurements on the CP violating phase $\phi_{s}$ and the mass
eigenstates width difference $\Delta \Gamma_{s}$ using $B_{s} \rightarrow J/\psi \phi$
decays.  The latest results have significantly reduced uncertainties and are
consistent with previous results.  Both experiments are planning on updating
their measurements using improved analysis techniques with the full data set.

%\Acknowledgements
%This work was supported by My Funding Agency under contract 12345. 
%I am grateful to M.Y. Colleague for useful discussions and suggestions.

\end{document}

%% file: econfmacros.tex
\def\beq{\begin{equation}}
\def\eeq#1{\label{#1}\end{equation}}
\def\eeqn{\end{equation}}

\def\beqa{\begin{eqnarray}}
\def\eeqa#1{\label{#1}\end{eqnarray}}
\def\eeqan{\end{eqnarray}}

\let\bar=\overbar

\def\Dslash{\not{\hbox{\kern-4pt $D$}}}
\def\dslash{\not{\hbox{\kern-2pt $\del$}}}

\def\msb{{\bar{\ssstyle M \kern -1pt S}}}